\documentclass[twocolumn,preprintnumbers,
               superscriptaddress,
               eqsecnum,nofootinbib,
               prd]{revtex4-1}
\usepackage{graphicx, fancybox}
\usepackage{amsmath,amssymb}
\usepackage[colorlinks=true, pdfstartview=FitV, linkcolor=red, citecolor=blue, urlcolor=blue]{hyperref}

\usepackage{slashed}
\newcommand{\sla}{\slashed}
\newcommand{\nn}{\nonumber}
\newcommand{\prj}{{\mathcal P}}
\newcommand{\gam}{\gamma}
\newcommand{\para}{\parallel}
\newcommand{\Ham}{ {\cal H} }

\newcommand{\vect}[1]{\mathbf{#1}}

\newcommand{\sgn}{\mathrm{sgn}}
\newcommand{\Slash}[1]{\ooalign{\hfil/\hfil\crcr$#1$}}

\newcommand{\vp}{\vect{p}}

\newcommand{\vzero}{\vect{0}}

\newcommand{\Bf}{B_f}
\newcommand{\eL}{\epsilon^L}

\newcommand{\qf}{q_f}
\newcommand{\expansion}{
\theta
}

\newcommand{\cond}{\kappa} 

\newcommand{\Tr}{\mathrm{Tr}}

\newcommand{\comment}[1]{}

\newcommand{\Tc}{{\mathrm T}_C}

\newcommand{\nf}{n_F}
\newcommand{\nb}{n_B}
\newcommand{\Nc}{N_c}
\newcommand{\Nf}{N_f}
\newcommand{\mf}{m_f}
\newcommand{\Cf}{C_f}

\usepackage{color}

\begin{document}

\title{
Bulk Viscosity
of Quark-Gluon Plasma in Strong Magnetic Fields
}

\author{Koichi Hattori}
\email{koichi.hattori@outlook.com}
\affiliation{Physics Department and Center for Particle Physics and Field Theory,
Fudan University, Shanghai 200433, China}

\author{Xu-Guang Huang}
\affiliation{Physics Department and Center for Particle Physics and Field Theory,
Fudan University, Shanghai 200433, China}
\affiliation{Key Laboratory of Nuclear Physics and Ion-beam Application (MOE), Fudan University, Shanghai 200433, China}

\author{Dirk H.\ Rischke}
\affiliation{Goethe University Frankfurt am Main, Institute for Theoretical Physics, Max-von-Laue-Str.\ 1, D-60438 Frankfurt am Main, Germany}
\affiliation{Department of
Modern Physics, University of Science and Technology of China, Hefei,
Anhui 230026, China}

\author{Daisuke Satow}
\email{dsato@th.physik.uni-frankfurt.de}
\affiliation{Goethe University Frankfurt am Main, Institute for Theoretical Physics, Max-von-Laue-Str.\ 1, D-60438 Frankfurt am Main, Germany}

\begin{abstract}
We investigate the viscosities of the quark-gluon plasma in strong magnetic fields
within the leading-log and lowest Landau level (LLL) approximations.
We first show that the bulk viscosity in the direction parallel to the magnetic field
is the only component that has a contribution from the quarks occupying the LLL.
We then compute the bulk viscosity from the Kubo formula
and find an intriguing quark-mass dependence as a consequence of a
competition between the suppression of the bulk viscosity by conformal symmetry
and an enhancement of the mean-free path by chirality conservation,
which governs the behavior in the massless limit.
The quark contribution to the viscosity along the magnetic field becomes larger than
the one in the absence of a magnetic field.
We also briefly estimate the other transport coefficients by considering the contribution of gluons.
We show that the shear viscosities are suppressed compared to their values in the absence of a magnetic field.
\end{abstract}

\date{\today}

\maketitle

\section{Introduction}
\label{sec:intro}

Heavy-ion collisions are the only way to experimentally investigate
the quark-gluon plasma (QGP), a form of matter composed of
quarks and gluons liberated from color confinement at high temperatures ($T$), under
controlled laboratory conditions.
At the same time, such experiments may provide us with an opportunity
to investigate QGP matter under the influence of strong magnetic fields ($B$),
since non-central heavy-ion collisions are thought to generate (via Ampere's law)
the strongest magnetic fields
ever created in terrestrial experiments~\cite{
Skokov:2009qp, *Voronyuk:2011jd, *Bzdak:2011yy, *Deng:2012pc, *Tuchin:2013apa, *Tuchin:2015oka,
*Holliday:2016lbx} [see Refs.~\cite{Huang:2015oca, Hattori:2016emy} for recent reviews].

Hydrodynamic simulations have played an important role
in the study of phenomenological aspects of heavy-ion collisions.
Recent efforts are directed towards applying magnetohydrodynamics (MHD),
which takes into account the dynamical coupling of the magnetic field to the fluid in a self-consistent 
way~\cite{Inghirami:2016iru, Pu:2016ayh, Roy:2015kma, Pu:2016bxy, Roy:2017yvg, Moghaddam:2017myy}.
This is an important progress in the investigation of the QGP in strong magnetic fields.
However, these studies have not yet implemented transport coefficients
computed in the presence of a magnetic field.

In the studies \cite{Hattori:2016lqx, Hattori:2016cnt},
the authors have computed the electrical conductivity
in a strong magnetic field. In this case, quarks are confined to the lowest Landau level (LLL).
It was shown that the microscopic properties of the LLL dynamics
manifest themselves as drastic modifications of the macroscopic transport properties.
The key observation was that there is a mismatch between the spatial dimensions
in which the quarks reside as compared to gluons:
LLL quarks can only propagate in one spatial dimension (parallel to the magnetic field),
while gluons can move in all three spatial dimensions.
This mismatch of dimensions opens a kinematical window
for 1-to-2 scattering~\cite{Hattori:2016lqx, Hattori:2016cnt,  Elmfors:1995gr},
and this emergent contribution dominates over the conventional
leading-order contributions from 2-to-2 scatterings at weak coupling $ (g \ll 1) $.
On the other hand, the chirality-conservation law for one-dimensional quarks
strictly prohibits scatterings in the massless limit ($ m_f =0 $).
Therefore, the parametric dependence of the quark damping rate
has been established as $ \sim g^2 m_f^2{/T} $,
up to a logarithmic factor [see Refs.~\cite{Hattori:2016lqx, Hattori:2016cnt} and Sec.~\ref{sec:bulk}],
which significantly enhances the electrical conductivity.
This dependence is one of the intriguing manifestations of the LLL dynamics,\footnote{
See also Refs.~\cite{Sadofyev:2015tmb, Fukushima:2015wck} for
a consequence of LLL kinematics, which manifests itself in the drag force.}
and serves as motivation to investigate other transport coefficients, in order to see whether LLL
dynamics has a similarly important influence.
Whereas LLL dynamics has been intensively investigated in studies of
anomalous transport phenomena [see, e.g.\ Refs.~\cite{
Kharzeev:2015znc, Huang:2015oca, Hattori:2016emy} for reviews],
its manifestation in the transport coefficients of MHD
has not been fully explored yet.

In this paper, we evaluate the {contribution of the LLL quarks to the} transport coefficients of MHD, 
on the basis of the aforementioned quark damping rate
and the Kubo formulas obtained in Ref.~\cite{Huang:2011dc}.
Since the magnetic field breaks the isotropy of the system,
in general MHD has more independent transport coefficients
than conventional (isotropic) hydrodynamics.
However, we will show that, within the LLL approximation, LLL quarks contribute
only to the component of the bulk viscosity parallel to the magnetic field.
[The only other transport coefficient with a contribution from LLL quarks is the longitudinal 
conductivity~\cite{Hattori:2016lqx, Hattori:2016cnt}.]
We find a nontrivial dependence of the bulk viscosity on the current quark mass as a result of
the competition between the chirality-conservation law and conformal symmetry.
To evaluate the bulk viscosity, we apply the same method that was used to
evaluate the electrical conductivity~\cite{Hattori:2016lqx, Hattori:2016cnt}.
The analyses of the present paper together with
those in Refs.~\cite{Hattori:2016lqx, Hattori:2016cnt} conclude
the computation of the LLL-quark contribution to
a certain set of transport coefficients shown in Ref.~\cite{Huang:2011dc},
within the leading-log and LLL approximations.
Other related works on the transport coefficients of MHD are, for example, the calculation of the 
shear viscosities in a weak magnetic field~\cite{Li:2017tgi, Tuchin:2011jw} and in the 
holographic setup~\cite{Rebhan:2011vd, Giataganas:2013hwa, Jain:2015txa, Finazzo:2016mhm}, 
and the calculation of the anisotropic bulk viscosities due to electroweak interactions in dense quark matter~\cite{Huang:2009ue}.

This paper is organized as follows:
In the next two sections, we recapitulate the basic equations of MHD and the Kubo formulas,
and then identify the relevant components of the viscosities.
In Sec.~\ref{sec:bulk}, we evaluate the component of the bulk viscosity parallel to the magnetic field
within the leading-log and LLL approximations.
Section~\ref{sec:implication} is devoted to 
evaluating the quark contribution to the bulk viscosity in heavy-ion collisions.
In Sec.~\ref{sec:gluon}, we make order-of-magnitude estimates for the gluon contribution to the shear and bulk viscosities.
We conclude with a summary of our results
in Sec.~\ref{sec:summary}.
In the first appendix, we briefly discuss the Landau-level quantization
and, in the other appendices,
evaluate the thermodynamic quantities in the LLL approximation
which are necessary for the evaluation of the bulk viscosity.
We also derive an expression for the bulk viscosity
from the linearized Boltzmann equation in (1+1) dimensions which is consistent with our result
obtained via the Kubo formula.

\section{Magnetohydrodynamics and Kubo formulas}

In this section, we briefly summarize the results of Ref.~\cite{Huang:2011dc},
which comprise the equations of motion of MHD and the constitutive relations.
In the presence of a magnetic field, MHD contains two bulk viscosities,
five shear viscosities, and three electrical conductivities.
We also recapitulate the Kubo formulas for these transport coefficients.

\subsection{Magnetohydrodynamics}
\label{ssc:MHD}

The basic equations of MHD consist of the conservation laws of energy, momentum,
and electric charge, and the constitutive equations
for the energy-momentum tensor ($T^{\mu \nu}$) and the electric-charge current  
($j^\mu$).\footnote{For the sake of simplicity, even in the case of
multiple flavors we consider only the electric-charge current.}
The former ones are given by
\begin{align}
\label{eq:hydroEOM-1}
\partial_\mu j^\mu
&=0,\\
\label{eq:hydroEOM-2}
\partial_\mu T^{\mu\nu}
&= F^{\nu\mu}j_{\mu},
\end{align}
where $F^{\mu\nu}$ is the electromagnetic field-strength tensor.
If the electric field is much smaller than the magnetic field,
the right-hand side of the second equation can be neglected~\cite{Huang:2011dc}.
For the evaluation of the transport coefficients, which is the purpose of this paper, 
it suffices to consider a static and homogeneous (non-dynamical) magnetic field.

The constitutive 
equations\footnote{In addition to the terms given in this expression, other terms are 
generated by the coupling between the vorticity and the magnetic field~\cite{Hernandez:2017mch}.
However, they are not subject of this paper, so we have omitted them.}
in the Landau frame read~\cite{Huang:2011dc}
\begin{align} 
j^\mu
&= n u^\mu+ {\cal J}^\mu,\\
T^{\mu\nu}
&= \epsilon u^\mu u^\nu -P_\perp \varXi^{\mu\nu} +P_\parallel b^\mu b^\nu
+{\cal T}^{\mu\nu},
\label{eq:constitutive-T}
\end{align}
where $u^\mu$ is the flow vector, normalized as $u^2=1$,
and  $b^\mu\equiv \epsilon^{\mu\nu\alpha\beta} F_{\nu\alpha} u_\beta/(2B) $
with $B\equiv \sqrt{-B^\mu B_\mu}$.
The tensor which projects onto the three-dimensional space orthogonal to the flow
is defined as $\Delta^{\mu\nu}\equiv g^{\mu\nu}-u^\mu u^\nu$, while that
which projects onto the two-dimensional space orthogonal to both the flow and the magnetic field
is $\varXi^{\mu\nu}\equiv \Delta^{\mu\nu}+b^\mu b^\nu$, respectively.
We also have the energy density $\epsilon$, the charge density $n$,
the thermodynamic pressure $P_\parallel\equiv P$,
and the transverse pressure $P_\perp\equiv P-MB$, including
the contribution of the magnetization $M\equiv (\partial P/\partial B)_{T,\mu}$
($\mu$ is the chemical potential associated with the electric charge).
To leading order of the derivative expansion the dissipative terms are given by
\begin{align}
\label{eq:jmu-dis}
{\cal{J}}^\mu
&= T\left(\cond_\perp \varXi^{\mu\nu} \nabla_\nu \alpha
-\cond_\parallel b^\mu b^\nu \nabla_\nu \alpha
-\cond_\times b^{\mu\nu} \nabla_\nu \alpha
\right) ,\\
\nonumber
{\cal T}^{\mu\nu}&= \frac{3}{2}\zeta_\perp \varXi^{\mu\nu} \phi
+3\zeta_\parallel b^\mu b^\nu \psi
+2\eta_0\left(w^{\mu\nu}-\frac{1}{3}\Delta^{\mu\nu}\theta\right) \\
\nonumber
&~~~+\eta_1 \left(\Delta^{\mu\nu}-\frac{3}{2}\varXi^{\mu\nu}\right)\left(\theta-\frac{3}{2}\phi\right) \\
\nonumber
&~~~+2\Bigl[-\eta_2\left(b^\mu\varXi^{\nu\alpha}b^\beta +b^\nu\varXi^{\mu\alpha}b^\beta \right) \\
\nonumber
&~~~-\eta_3 \left(\varXi^{\mu\alpha}b^{\nu\beta} +\varXi^{\nu\alpha}b^{\mu\beta} \right) \\
\label{eq:Tmunu-dis}
&~~~+\eta_4 \left(b^{\mu\alpha}b^{\nu}b^{\beta} +b^{\nu\alpha}b^{\mu}b^{\beta} \right)\Bigr]
w_{\alpha\beta},
\end{align}
where $\alpha\equiv\beta\mu$, $b^{\mu\nu}\equiv \epsilon^{\mu\nu\alpha\beta} b_\alpha u_\beta$, 
$w^{\mu\nu}\equiv (\nabla^\mu u^\nu+\nabla^\nu u^\mu)/2$, $\phi\equiv \varXi^{\mu\nu} w_{\mu\nu}$, 
$\psi\equiv b^\mu b^\nu w_{\mu\nu}$, $\theta\equiv \partial^\mu u_\mu$, 
with $\nabla_\mu\equiv \Delta _{\mu\nu} \partial^\nu$. 
The three $\cond$'s are the electrical conductivities, the two $\zeta$'s the bulk viscosities, and
the five $\eta$'s the shear viscosities, respectively.

\subsection{Kubo formulas}
\label{ssc:Kubo-formula}

The transport coefficients in Eqs.~(\ref{eq:jmu-dis}) and (\ref{eq:Tmunu-dis}) are given by the following 
Kubo formulas~\cite{Huang:2011dc}
\begin{align}
\label{eq:Kuboformula-kappa-para}
\cond_\parallel
&=  \frac{\partial}{\partial\omega} {\text{Im}}
G^R_{j^3 j^3}|_{\vp=\vzero, \omega\rightarrow 0} ,\\
\label{eq:Kuboformula-zeta-para}
\zeta_\parallel
&= \frac{1}{3} \frac{\partial}{\partial\omega}
\left(2 \text{Im}  G^R_{\tilde{P}_\perp \tilde{P}_\parallel }
+{\text{Im}} G^R_{\tilde{P}_\parallel\tilde{P}_\parallel}
\right)_{\vp=\vzero, \omega\rightarrow 0},\\
\zeta_\perp
&= \frac{1}{3} \frac{\partial}{\partial\omega}
\left( 2 {\text{Im}} G^R_{\tilde{P}_\perp \tilde{P}_\perp }
+ \text{Im}  G^R_{\tilde{P}_\parallel\tilde{P}_\perp}
\right)_{\vp=\vzero, \omega\rightarrow 0},\\
\eta_0
&=  \frac{\partial}{\partial\omega} {\text{Im}}
G^R_{T^{12}T^{12}}|_{\vp=\vzero, \omega\rightarrow 0}
,\\
\eta_1
&= -\frac{4}{3}\eta_0
-2 \frac{\partial}{\partial\omega}
\text{Im}  G^R_{\tilde{P}_\parallel \tilde{P}_\perp}|_{\vp=\vzero, \omega\rightarrow 0}
,\\
\eta_2
&=-\eta_0
+  \frac{\partial}{\partial\omega}
{\text{Im}} G^R_{T^{13}T^{13}}|_{\vp=\vzero, \omega\rightarrow 0}
,\\
\eta_3
&= \frac{1}{2} \frac{\partial}{\partial\omega}
\text{Im} G^R_{\tilde{P}_\perp T^{12}}|_{\vp=\vzero, \omega\rightarrow 0} ,\\
\label{eq:Kuboformula-eta4}
\eta_4
&= \frac{\partial}{\partial\omega}
\text{Im}  G^R_{T^{13}T^{23}}|_{\vp=\vzero, \omega\rightarrow 0},
\end{align}
where  without loss of generality the direction of the magnetic field is chosen to point along the 3-direction.
We have defined $\tilde{P}_\parallel\equiv P_\parallel -\Theta_\beta \epsilon$,
$\tilde{P}_\perp\equiv P_\perp -(\Theta_\beta +\Phi_\beta) \epsilon$,
with $\Theta_\beta\equiv (\partial P_\parallel/\partial \epsilon)_{B}$
and $\Phi_\beta\equiv -B(\partial M/\partial \epsilon)_{B}$.
Here, the retarded Green's function is defined\footnote{
Note that this definition has the opposite sign compared to that in Ref.~\cite{Huang:2011dc}.
Because of this difference, the signs in Eqs.~(\ref{eq:Kuboformula-kappa-para})--(\ref{eq:Kuboformula-eta4}) are also opposite to those in Ref.~\cite{Huang:2011dc}.}
as $G^R_{AB}(x)\equiv i\theta(x^0)\langle [ A(x),B(0) ]\rangle $
with the average in the equilibrium state denoted by angular brackets.
Since we focus on the charge-neutral case in this paper,
the above Kubo formulas lack some terms that are
present in the nonzero-charge case~\cite{Huang:2011dc}.

We also note that the right-hand side of Eq.~(\ref{eq:hydroEOM-2}) should be maintained
to derive the correct Kubo formulas for $\cond_\perp$ and $\cond_\times$.
However, since they vanish in the LLL approximation, we do not go into this issue in the present analysis.
For a discussion of the complete Kubo formulas, see Ref.~\cite{Hernandez:2017mch}.


\section{Contributions of the LLL quarks to transport coefficients}
\label{sec:LLL}

Before we compute the transport coefficients in the subsequent sections,
we briefly describe the LLL approximation and identify the transport coefficients
which have contributions from the quarks in the LLL.

The periodic cyclotron motion in a magnetic field leads to the Landau-level quantization,
the energies of which are 
\begin{eqnarray}
\epsilon_n = \sqrt{ (p^3)^2 + m_f^2 + 2 n |B_f|}
\label{eq:disp}
\, ,
\end{eqnarray}
specified by a non-negative integer ($ n\geq0 $).
We have defined $ \Bf = |e|\qf B $ with $  |e| q_f$ being the electric charge,
and we explicitly maintain the current quark mass $ m_f $ as its dependence turns out to be important later.
In this paper, we focus on the strong-field regime that
satisfies the hierarchy $ |B_f| \gg T^2 \gg m_f^2 $.
Therefore, the occupation number of quarks in the LLL ($  n=0$),  the energy of which is 
independent of $ |B_f| $, is large in an ensemble at temperature $T$.
On the other hand, the occupation number of the higher Landau levels (hLLs) [$n\geq 1$], 
which are separated from the LLL by
a large energy gap of the order of $ \sqrt{ |B_f| } $, is highly suppressed by the Boltzmann factor.
Thus, we entirely neglect the hLLs in our calculation.

We work in the Landau gauge specified as
$A^{\text{ext}}_2=Bx_1$, with the other components vanishing.
Therefore, three components of the quark momentum,
which is denoted as $ \bar p^\mu = (p^0,0,p^2,p^3) $, are still good quantum numbers in a magnetic field.
The final one is provided by the principle quantum number $ n $.
The energy eigenstates specified by these quantum numbers are complete and orthogonal.
Therefore, by using the eigenfunction shown in Appendix~\ref{app:Dirac-eq},
the quark field can be expanded as~\cite{Hattori:2016cnt, Hattori:2015aki}
\begin{align}
\label{eq:expansion}
\psi(x)
&=
\int _{\bar p} e^{- i \bar p \cdot x} {\cal H}(\tilde x^f_p)
{\cal P}_+ \, \chi (p_L)
\\
& \hspace{0.5cm}
+ \ ({\rm Contributions \ from \ {\it n}\geq1})
\nn
\, ,
\end{align}
where the explicit expression of the contribution from
the hLLs was suppressed as it will not be discussed below.
We have introduced the abbreviations
$\int_p\equiv \int dp/(2\pi)$, $p_L^\mu \equiv (p^0,0,0,p^3)$, and
$ \tilde x^f_p \equiv x^1- p^2/B_f $ with
the second component of the momentum, $ p^2 $.
The spin projection operator is defined by
${\cal P}_\pm\equiv [1\pm \sgn(\Bf)i\gamma^1\gamma^2]/2$ with a sign function $\sgn(\Bf)$.
The spin of the LLL quark is frozen in a definite direction
along the magnetic field due to the Zeeman effect.
${\cal H}(x)$ is the normalized Hermite function
coming from the quark wave function in the transverse plane in the LLL.
In Appendix~\ref{app:Dirac-eq}, we summarize the properties of ${\cal H}(x)$
which will be used below.

The spin projection operator has the useful property
$  \prj_\pm \gam^\mu \prj _\pm = \gam_L^\mu \prj_\pm$, with $\gam_L^\mu = (\gam^0,0,0,\gam^3)  $.
Therefore, the current composed of the LLL quark field is
\begin{eqnarray}
j^\mu (x) = \bar \psi (x) \gam_L^\mu \psi (x)
\, ,
\end{eqnarray}
which has only a temporal as well as a spatial component parallel to the magnetic field.
The transverse components demand a spin flip which, however, costs energy
for an inter-level transition from $ n=0 $ to $ n\geq 1 $.
This is the reason why the LLL quarks contribute only to the longitudinal component
of the conductivity shown in Eq.~(\ref{eq:Kuboformula-kappa-para}).
The computation of the longitudinal conductivity
has been performed in Refs.~\cite{Hattori:2016lqx, Hattori:2016cnt}.

Similar to the current, we now identify
the nonvanishing components of the energy-momentum tensor in the LLL
and the relevant contributions to the viscosities listed
in Eqs.~(\ref{eq:Kuboformula-zeta-para})--(\ref{eq:Kuboformula-eta4}).
In the absence of QCD interactions, the quark part of
the energy-momentum tensor is written as\footnote{
The trace part vanishes if one uses the equation of motion, which is the on-shell condition for the quark.
In our calculation, this condition applies, so we do not write the trace part here.}
\begin{align}
\label{eq:e-m}
T^{\mu\nu}(x)
&=\frac{i}{2} {\cal S} \sum_f \left[ \overline{\psi} \overleftarrow{D}^\mu \gamma^\nu\psi
+\overline{\psi} D^\mu\gamma^\nu \psi
\right],
\end{align}
where the sum is taken over the flavor index.
The covariant derivatives with the external magnetic field are
$D_\mu\equiv \partial_\mu+i q_f |e| A^{\text{ext}}_\mu$ and 
$\overleftarrow{D}_\mu\equiv -\overleftarrow{\partial}_\mu+i q_f |e| A^{\text{ext}}_\mu$.
The symmetrization operator works as ${\cal S} f^{\mu\nu}\equiv (f^{\mu\nu}+f^{\nu\mu})/2$.

Consider one of the four terms in Eq.~(\ref{eq:e-m}), e.g.
\begin{eqnarray}
t^{\mu\nu} (q_L) \equiv
\frac{i}{2} \int \!\! d^4x \, e^{-iq_L\cdot x_L} \bar \psi (x)  \, \gam^\mu D^\nu \, \psi(x)
\label{eq:t1}
\, .
\end{eqnarray}
Since we are interested in the transport coefficients in the static and homogeneous limit,
we have taken the transverse momentum to be zero, $ q_{1,2} = 0 $.
(For the moment, we shall keep a finite $ q_3 $ for notational simplicity.)
Inserting the expansion of the quark field (\ref{eq:expansion}) into the above term, we have
\begin{eqnarray}
\label{eq:t2}
t^{\mu\nu} (q_L) = \frac{1}{2} \int _{\bar p}
\bar \chi (p_L +q_L)   \gam^\mu_L  \Gamma^\nu (\bar p) \prj_+ \chi ( p_L )
,
\end{eqnarray}
where
\begin{eqnarray}
\label{eq:Gam1}
\Gamma^\nu (\bar p) =  \int \!\! d x^1 \,
\Ham ( \tilde x^f_p ) ( \, p^\nu_L -  \delta^{\nu 2}  B_f \tilde x^f_p
 + i \delta^{\nu 1}  \partial^1 \,  ) \Ham( \tilde x_p^f )
 \, .  \nn \\
\end{eqnarray}
According to the integral formulas given in Appendix~\ref{app:Dirac-eq},
only the first term survives:
\begin{eqnarray}
\label{eq:Gam2}
\Gamma^\nu (\bar p) &=&   p_L^\nu
\, ,
\end{eqnarray}
which does not depend on the second component $p_2  $ within the LLL approximation.
Then, we find
\begin{eqnarray}
t^{\mu\nu} (q_L)
=  \frac{1}{2} \int_{\bar p}
\bar \chi (p_L + q_L )  \gam^\mu_L  p_L^\nu \prj_+  \chi (p_L )
\, .
\end{eqnarray}
An important point is that $ t^{\mu\nu} $ has nonvanishing entries
only in the longitudinal components specified by $\mu,\,\nu = 0,\,3  $,
and vanishes when either one, or both, of the indices are $1$ or $  2$.
The same conclusion is drawn for the other three terms of
the energy-momentum tensor (\ref{eq:e-m}).
This is a natural consequence of the fact that
the LLL quarks can carry only an energy and a momentum
parallel to the magnetic field.\footnote{
This conclusion could be modified
when the hLLs contribute to the energy-momentum tensor or
when the external transverse momentum $ q_\perp $ (spatial modulation) is finite.
}

From the above observation, we find that
only the longitudinal pressure $ \tilde P_\para   $ has contributions from the LLL quarks
among all Eqs.~(\ref{eq:Kuboformula-zeta-para})--(\ref{eq:Kuboformula-eta4}).
Therefore, the LLL quark carriers mainly contribute to
the longitudinal component of the bulk viscosity $ \zeta_\para $
which is given by the correlator of the diagonal components,
$ \tilde P_\para = T^{\mu\nu} b_\mu b_\nu - \Theta_\beta T^{\mu\nu} u_\mu u_\nu
=  T^{33}  - \Theta_\beta T^{00} $.
To reach the above conclusion, note also that, according to Eq.~(\ref{eq:Phibeta-Thetabeta}),
we have $\tilde P_\perp = P_\perp  $ in the LLL which is solely given by
the transverse components of the energy-momentum tensor.

All other correlators in Eq.~(\ref{eq:Kuboformula-zeta-para}) also have contributions from gluons, 
of which we will make order-of-magnitude estimates in Sec.~\ref{sec:gluon}.

\section{Bulk viscosity}
\label{sec:bulk}

Having identified the contribution of the LLL quarks to the viscosity,
we now evaluate the relevant component,
that is, the longitudinal component of the bulk viscosity ($\zeta_\parallel$).
We also discuss the physical meaning of the result.

\subsection{Calculation}

\begin{figure}[t!]
\begin{center}
\includegraphics[width=0.3\textwidth]{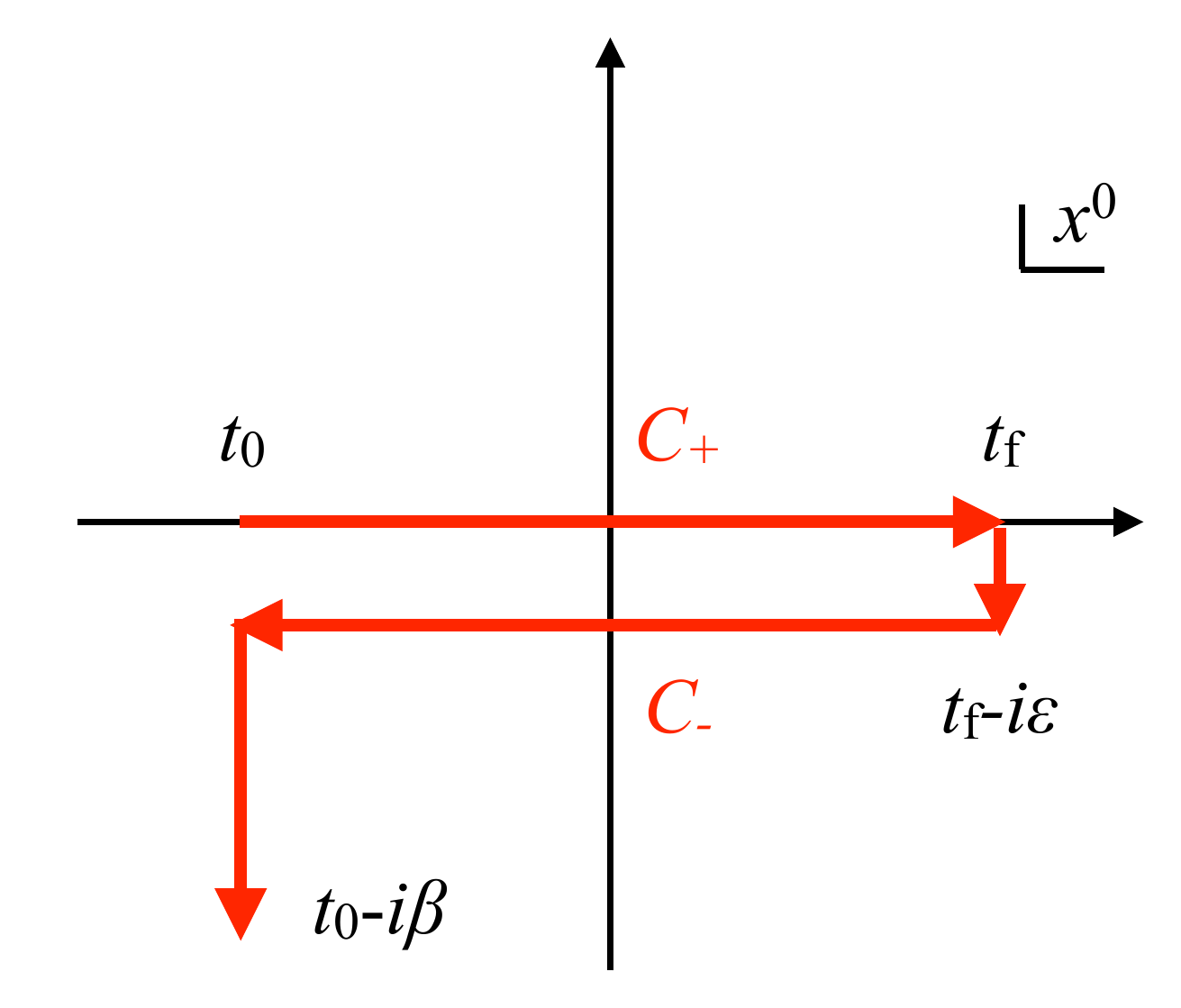}
\caption{The contour in the complex-time plane.
The part $C_+$ runs along the real axis and the part $C_-$ runs parallel to this axis, but is
displaced by $-i\varepsilon$.
}
\label{fig:contour}
\end{center}
\end{figure}

We use the real-time formalism~\cite{Bellac:2011kqa, Blaizot:2001nr}
for the diagrammatic calculation of the bulk viscosity.
First, by using the (12) basis,
we write the retarded Green's function in Eq.~(\ref{eq:Kuboformula-zeta-para}) in the low-energy limit as
\begin{align}
\label{eq:GR-Pperp-G12}
 \frac{\partial}{\partial\omega}
 {\text{Im}} G^R_{\tilde{P}_\parallel\tilde{P}_\parallel}(\omega,\vzero)
&\simeq \frac{\beta}{2} G^{(12)}_{\tilde{P}_\parallel\tilde{P}_\parallel}(p=0),
\end{align}
where $G^{(12)}_{AB}\equiv \langle \Tc A_1(x)B_2(0)\rangle= \langle B_2(0) A_1(x)\rangle$,
and $\Tc$ is the path-ordering operator on the complex-time path $C$,
which is plotted in Fig.~\ref{fig:contour}.
The operator with the index 1 (2) is defined on the path $C_+$ ($C_-$).
As discussed in the previous section, the transverse pressure $  \tilde{P}_\perp$
does not have a contribution from the LLL quarks,
so that the other correlator, $G^R_{\tilde{P}_\perp \tilde{P}_\parallel }$,
in Eq.~(\ref{eq:Kuboformula-zeta-para}) is much smaller than the one in Eq.~(\ref{eq:GR-Pperp-G12}).
Therefore, we neglect this contribution in the following.

From Eq.~(\ref{eq:constitutive-T}), the relevant pressure component is expressed
as $  \tilde P_\para(p) = T^{33} (p)  - \Theta_\beta T^{00}(p)$.
Thus, the Green's function with distinct external momenta reads
\begin{eqnarray}
\begin{split}
G^{(12)}_{\tilde{P}_{\parallel} \tilde{P}_{\parallel}}(p, p^\prime)
&= \big \langle \, {\Tc} \tilde P_\para(p) \tilde P_\para(p^\prime) \, \big\rangle
\\
&= \int_{\bar k} \int_{\bar k^\prime}
\big\langle \, \bar \chi_{ 2}(p_L+k_L) \sla k_\Theta \prj_+ \chi_{ 2}(k)
\\
& \hspace{1.5cm} \times
\bar \chi_{ 1}(p_L^\prime +k_L^\prime ) \sla k_\Theta^\prime \prj_+ \chi_{ 1}(k^\prime)  \, \big\rangle
\label{eq:Green1}
 ,
 \end{split}
\end{eqnarray}
where we have defined $ \sla k_\Theta \equiv  k^3 \gam^3 - \Theta_\beta k^0 \gam^0  $
and inserted the expression of the energy-momentum tensor
for vanishing transverse momentum $ p_\perp = p_\perp^\prime =0 $
which was discussed in the previous section.
All four of the terms in Eq.~(\ref{eq:e-m}) result in the same expression
up to differences which will vanish in the end when we take the limit $ p, p^\prime \to 0 $.

\begin{figure}[t!]
\begin{center}
\includegraphics[width=0.2\textwidth]{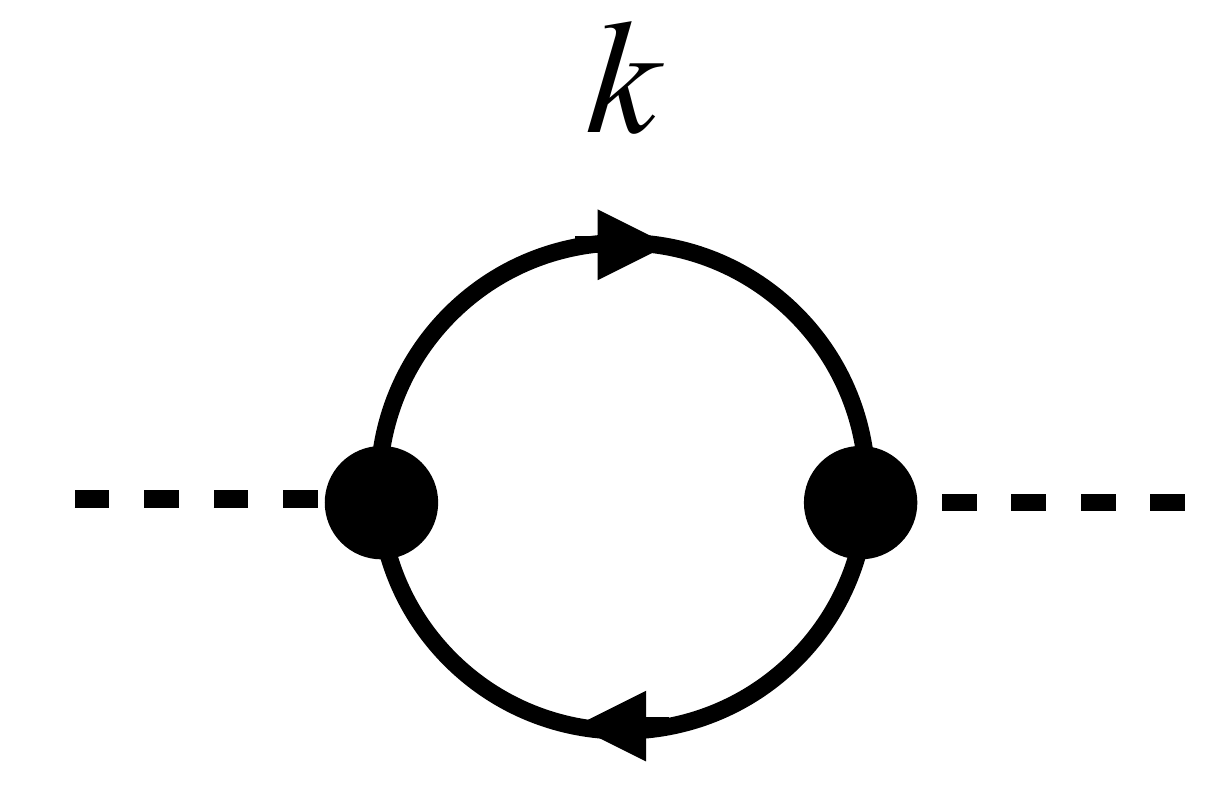}
\caption{The one-loop diagram for $G^{(12)}_{\tilde{P}_\parallel\tilde{P}_\parallel}$.
The solid line is the quark propagator, and the blob is the vertex for $\tilde{P}_\parallel$, which is 
$k^3\gamma^3-\Theta_\beta k^0\gamma^0$.
}
\label{fig:one-loop}
\end{center}
\end{figure}

We start with the one-loop approximation.
In this approximation, we can evaluate the Green's function by using Wick's theorem (the corresponding diagram is drawn in Fig.~\ref{fig:one-loop}).
The thermal averages of the spinors are replaced by the thermal LLL propagator
$S^{(ij)}(k_L, k^\prime_L)
=  \langle\chi_i (k_L) \overline{\chi}_j (k_L^\prime)  \rangle
= \delta^{(3)}(\bar k - \bar k^\prime) S^{(ij)}(k_L) $.
Note that there is three-dimensional momentum conservation in the Landau gauge.
Inserting the propagators into the Green's function,
we will, therefore, get a delta function, $   \delta^{(3)} (p+p^\prime)$,
for the overall conservation of external momenta, 
which in turn becomes the (three-dimensional) system volume in the limit $ p,p^\prime \to 0 $.
Dividing the Green's function (\ref{eq:Green1}) by the volume $ V_4 = L_1 \delta^{(3)}(0) $
with $ L_1 $ being the length in the residual dimension, we have
\begin{eqnarray}
G^{(12)}_{\tilde{P}_\parallel\tilde{P}_\parallel}(p = 0)
= \frac{1}{V_4} G^{(12)}_{\tilde{P}_\parallel\tilde{P}_\parallel}(p=0 , p^\prime = 0)
\, ,
\end{eqnarray}
and the one-loop expression is found to be
\begin{align}
\label{eq:G12-oneloop}
\begin{split}
&
G^{(12)}_{\tilde{P}_\parallel\tilde{P}_\parallel}(p=0)
\\
&=- \Nc \sum_{f} \frac{|\Bf|}{2\pi}  \int_{k_L}
\Tr[ S^{(21)}(k_L) \sla k_{\Theta} {\cal P}_+ S^{(12)}(k_L)  \sla k_{\Theta} ] \\
 &= \Nc \sum_{f} \frac{|\Bf|}{2\pi} \int_{k_L}
\Tr\left[\left\{(\Slash{k}_L+\mf) \sla k_{\Theta} \right\}^2   {\cal P}_+ \right]
 \\
& \hspace{2.5cm} \times
\nf(k^0) [1-\nf(k^0)]  [\rho^S(k_L)]^2  .
\end{split}
\end{align}
In the first line, we obtained the density of states $ |B_f|/(2\pi) $ in the transverse plane
as explained in Appendix~\ref{app:Dirac-eq}.
In the last line, we have inserted $S^{(12)}(k_L)= -(\Slash{k}_L+\mf) \nf(k^0)\rho^S(k_L)$
and $S^{(21)}(k_L)= (\Slash{k}_L+\mf) [1-\nf(k^0)]\rho^S(k_L)$
with the quark spectral function $\rho^S(k_L)$
and the Fermi distribution function $\nf(k^0)\equiv[e^{\beta k^0}+1]^{-1}$.

The trace in Eq.~(\ref{eq:G12-oneloop}) is evaluated as
\begin{align}
\label{eq:trace}
\begin{split}
&\Tr[\left\{(\Slash{k}_L+\mf) \sla k_\Theta \right\}^2   {\cal P}_+ ]
= 4[(\eL_k)^2 X-\mf^2]^2,
\end{split}
\end{align}
where we have introduced $X=1-\Theta_\beta$, and used the on-shell condition 
$k^0=\pm \eL_k$ $(\eL_k\equiv \sqrt{(k^3)^2+\mf^2})$, which will be justified later.
According to Appendix~\ref{app:thermodynamic}, $X$ vanishes at $\mf=0$,
and so does the trace in the massless limit.

The square of the spectral function needs to be treated with care:
If we naively use the non-interacting form for the spectral function,
$\rho^S_0(k_L)=2\pi \, \sgn(k^0)\delta(k^2_L-\mf^2)$,
it diverges on account of a pinch singularity.
Physically, the viscosity is indeed expected to diverge in a free theory,
and becomes finite due to interactions.
Therefore, the spectral function resummed with a finite damping rate ($\xi_k$)
gives a finite result~\cite{Jeon:1994if, Gagnon:2007qt, Gagnon:2006hi, Hidaka:2010gh}.
By using $\rho^S(k_L)= 4\xi_k k^0/[(k^2_L-\mf^2)^2+(2\xi_k k^0)^2]$,
it is approximated as~\cite{Hattori:2016cnt, Wang:2002nba}
\begin{align}
\label{eq:spectral-square}
[\rho^S(k_L)]^2
\simeq
\frac{\rho^S_0(k_L)}{2\xi_k k^0},
\end{align}
where we have neglected terms that vanish after the $k^0$ integration.

The spectral function (\ref{eq:spectral-square}) and the trace in Eq.~(\ref{eq:trace})
allow us to express the Green's function (\ref{eq:G12-oneloop}) 
in terms of the damping rate.
Plugging these expressions into the Kubo formula (\ref{eq:Kuboformula-zeta-para}),
the bulk viscosity can be written as
\begin{align}
\label{eq:bulkvis-quarkdamp}
\begin{split}
\zeta_\parallel
&=  \frac{2\beta}{3}  \Nc\sum_f \frac{|\Bf|}{2\pi} \mf^4
\int_{k_L}  \left[\frac{3}{\pi^2T^2}(\eL_k)^2-1\right]^2 \\
&~~~\times \frac{\rho^S_0(k_L)}{2\xi_k k^0}  [1-\nf(k^0)] \nf(k^0),
\end{split}
\end{align}
where the expression of $X$ in Eq.~(\ref{eq:Theta-beta}) has been used.

The quark damping rate arising from 1-to-2 scattering
was already evaluated at the leading-log accuracy in Ref.~\cite{Hattori:2016cnt} in the following two cases:
One is the case $M_g\ll \mf\ll T$, with $M_g$ being the Schwinger mass of
the gluon~\cite{Fukushima:2011nu, Hattori:2017xoo},
\begin{align}
M_g^2
&\equiv
\frac{1}{2}\frac{g^2}{\pi} \sum_f \frac{|\Bf|}{2\pi},
\end{align}
where $g$ is the QCD coupling constant.
The result in this case reads
\begin{align}
\label{eq:damping-M<<m}
\epsilon^L_k \xi_k
&\simeq
\frac{g^2\Cf \mf^2}{4\pi}
\left[\frac{1}{2}+\nb(\eL_k)\right]
\ln\left(\frac{T}{\mf}\right),
\end{align}
at the leading-log accuracy, with $\Cf\equiv (\Nc^2-1)/(2\Nc)$ and $\nb(k^0)\equiv [e^{\beta k^0}-1]^{-1}$.
The factor of $\mf^2$ can be explained in terms of chirality conservation~\cite{Hattori:2016cnt, Smilga:1991xa}.
In the other case, we have $\mf \ll M_g \ll T$.
The expression in this case can be obtained by replacing the log in Eq.~(\ref{eq:damping-M<<m}) with $\ln(T/M_g)$.


\subsection{Results}

By combining Eqs.~(\ref{eq:bulkvis-quarkdamp}) and (\ref{eq:damping-M<<m}), we can evaluate $\zeta_\parallel$.
In the case $M_g\ll \mf$ the result reads
\begin{align}
\label{eq:result-bulkvis}
\begin{split}
\zeta_\parallel
&= \frac{8\beta}{3}  \Nc\sum_f \frac{|\Bf|}{2\pi} \frac{\mf^2}{g^2\Cf \ln\left(T/\mf \right)} \\
&~~~\times \int^\infty_0 dk^3  \frac{1}{\eL_k}
 \left[\frac{3}{\pi^2T^2}(\eL_k)^2-1\right]^2
\bar N(\eL_k) 
 \\
 &\simeq    \Nc \sum_f \frac{|\Bf|}{2\pi} \frac{\mf^2}{g^2\Cf T \ln\left(T/\mf \right)}
 \left[\frac{4}{\pi^2}
-\frac{56}{3}\zeta'(-2)\right] \\
&\sim eB T^2 \left(\frac{\mf^2}{T^2}\right)^2 \frac{T}{g^2\mf^2 \ln(T/\mf)},
\end{split}
\end{align}
where we have defined $ \bar N(\eL_k) \equiv  [1-\nf(\eL_k)] \nf^2(\eL_k)/ \nb(\eL_k)  $
and assumed $\mf\ll T$.
Useful integration formulas are given by
$\int^\infty_0 d\epsilon \epsilon^3 \bar N(\epsilon)=T^4\pi^2/2$,
$\int^\infty_0 d\epsilon \, \epsilon \, \bar N(\epsilon)=T^2/2$,
and $\int^\infty_0 d\epsilon \, \epsilon^{-1} \bar N(\eL_k)
= -7\zeta'(-2) $
with the first derivative of the zeta function $\zeta'(-2) \simeq -0.0304$.
Expression (\ref{eq:result-bulkvis}) is one of the central results of our paper, 
which is valid when the thermal excitations are well activated ($m_f \ll T  $) 
and the LLL approximation works ($ T \ll \sqrt{|B_f|}  $).\footnote{ 
When $m_f \ll M_g  $, the log factor is replaced by $\ln(T/M_g)  $ as indicated in the previous section. 
The former condition is also replaced by $M_g \ll T  $, accordingly.
}
Although the above result was obtained by the diagrammatic method 
applied to a Kubo formula, we can obtain the same result by 
using the Boltzmann equation in (1+1) dimensions, as is shown in Appendix~\ref{app:Boltzmann}.
We note that this quantity is proportional to $\mf^2$, so the $s$ quark contributes more than the $u$ and $d$ quarks.
This is in striking contrast to the longitudinal component of the electrical 
conductivity~\cite{Hattori:2016cnt, Hattori:2016lqx}, which is proportional to $\mf^{-2}$, and thus 
the $u$ and $d$ quarks dominate over the contribution from the $s$ quark.

Several remarks on the parametric behavior are in order:
To this end, we first recapitulate the behavior at $B=0$~\cite{Arnold:2006fz}, which reads
\begin{align}
\label{eq:zeta-B=0-schematic}
\begin{split}
\zeta_{B=0} &\sim
{\text{(typical momentum)}^4}
\frac{\text{(conformal breaking factor)}^2} {\text{(mean free path)}^{-1} } \\
&\sim T^4 \left( \frac{\mf^2}{T^2}\right)^2\frac{1}{g^4T\ln(1/g)}.
\end{split}
\end{align}
The physical origin of the two conformal breaking factors are clear in the calculation with the Boltzmann equation~\cite{Arnold:2006fz}, which is done in Appendix~\ref{app:Boltzmann}.
In comparison, we find the following points in our results:
\begin{itemize}
\item An overall factor of $|\Bf|$ appears. It originates from
the Landau degeneracy of the quarks in the transverse plane, which are carriers of the pressure.
This factor replaces one factor of $T^2$ in the expression for $B=0$, 
resulting in an enhancement of the bulk viscosity in a strong magnetic field, $|B_f| \gg T^2 $.

\item Since the dominant scattering process is 1-to-2~\cite{Elmfors:1995gr, Hattori:2016cnt, Hattori:2016lqx} 
instead of 2-to-2, the $g$ dependence of the quark damping rate ($\sim$ inverse of the mean free path) 
is $\sim g^2$, not $\sim g^4$.
Therefore, the denominator of the bulk viscosity is proportional to $g^{2}$.
\item The $\mf$ dependence is also different because of chirality conservation.
It yields a factor of $\mf^2/T$ in the inverse of the mean-free path,
and this factor partially cancels the $\mf$ dependence coming from the two conformal breaking factors $(\mf^2/T^2)^2$ in the numerator.
Thus, the $\mf$ dependence of $\zeta_\parallel$ becomes a quadratic one in the end,
meaning that the bulk viscosity decreases with a decreasing mass
more slowly than the one computed without the effects of the strong magnetic field.
This is a consequence of the competition between the two constraints
which govern the behavior in the massless limit.
\end{itemize}
These points are shown clearly in the final lines of Eqs.~(\ref{eq:result-bulkvis}) and (\ref{eq:zeta-B=0-schematic}).

Finally, we comment on the effect of higher-loop orders.
In general, ladder-diagram contributions could be of
the same order of magnitude as the one-loop diagram
when a pinch singularity appears~\cite{Jeon:1994if, Gagnon:2007qt, Gagnon:2006hi, Hidaka:2010gh}.
For this reason, one may wonder if one needs to resum all-order ladder diagrams
to obtain the correct leading-order result.
However, it is easy to show that the resummation is not required at the leading-log accuracy, 
meaning that our result (\ref{eq:result-bulkvis}) is correct at this order.

The proof follows the line shown in Sec.~5 of Ref.~\cite{Hattori:2016cnt}. 
There are two steps in the proof. 
The first step is to identify the terms which have the maximum number of pinch singularities 
and are potentially as large in magnitude as the one-loop contribution. 
As in the case without external magnetic field, 
one indeed finds, by using the r/a basis, the terms in which 
all pairs of fermion propagators facing each other 
in the ladder diagram have pinch singularities. 
In the second step, one obtains the Bethe-Salpeter equation for the ladder resummation, 
which results in a gauge-invariant integral equation. 
Inserting the explicit forms of the quark and gluon spectral functions, 
one finds that the iterative correction vanishes\footnote{
Nevertheless, the quark damping rate is computed within the leading-log approximation.
} 
when $ m_f \gg M_g $ 
and is suppressed by an inverse log factor $ 1/\ln(T/M_g) $ when $m_f \ll M_g   $. 
Therefore, in both cases the one-loop result is correct within the leading-log approximation, 
which is valid when the inverse log factor is small, i.e., 
when $  T/m_f \gg 1 $ and $ T/M_g \gg1 $, respectively. 

\section{Estimate of the bulk viscosity}
\label{sec:implication}

In this section, we compute the value of $\zeta_\parallel$ for some choices of the parameters which 
appear to be realistic for heavy-ion collisions, and compare it with the value at $B=0$.
We use the following values for the parameters:
\begin{align}
\label{eq:parameter}
\begin{split}
\alpha_s &\equiv \frac{g^2}{4\pi}=0.3 ,\\
\Nc&=3,\\
eB &= 10 m^2_\pi =(443 {\text{ MeV}})^2, \\
\mf &= 100{\text{ MeV}}~~(s{\text{ quark}}),
\end{split}
\end{align}
where we have assumed a strong magnetic field, with $m_\pi = 140$ MeV being the pion mass.
For the case $\Nf=3$, the parameters above yield $M_g\simeq 160$ MeV.
Because it is larger than $\mf$, we use the expression for the bulk viscosity in the case $M_g\gg \mf$:
\begin{align}
\label{eq:zeta1}
\begin{split}
\zeta_\parallel
 &=  \Nc\frac{|\Bf|}{2\pi} \frac{\mf^2}{g^2\Cf T \ln\left(T/M_g \right)}
 \left[\frac{4}{\pi^2}
-\frac{56}{3}\zeta'(-2)\right]\\
&\simeq  0.031\frac{|eB|\mf^2}{T \ln\left(T/M_g \right)},
\end{split}
\end{align}
where we have taken only the contribution from the $s$ quark, since it is dominating over the contributions 
from the other flavors.

\begin{figure}[t!]
\begin{center}
\includegraphics[width=0.5\textwidth]{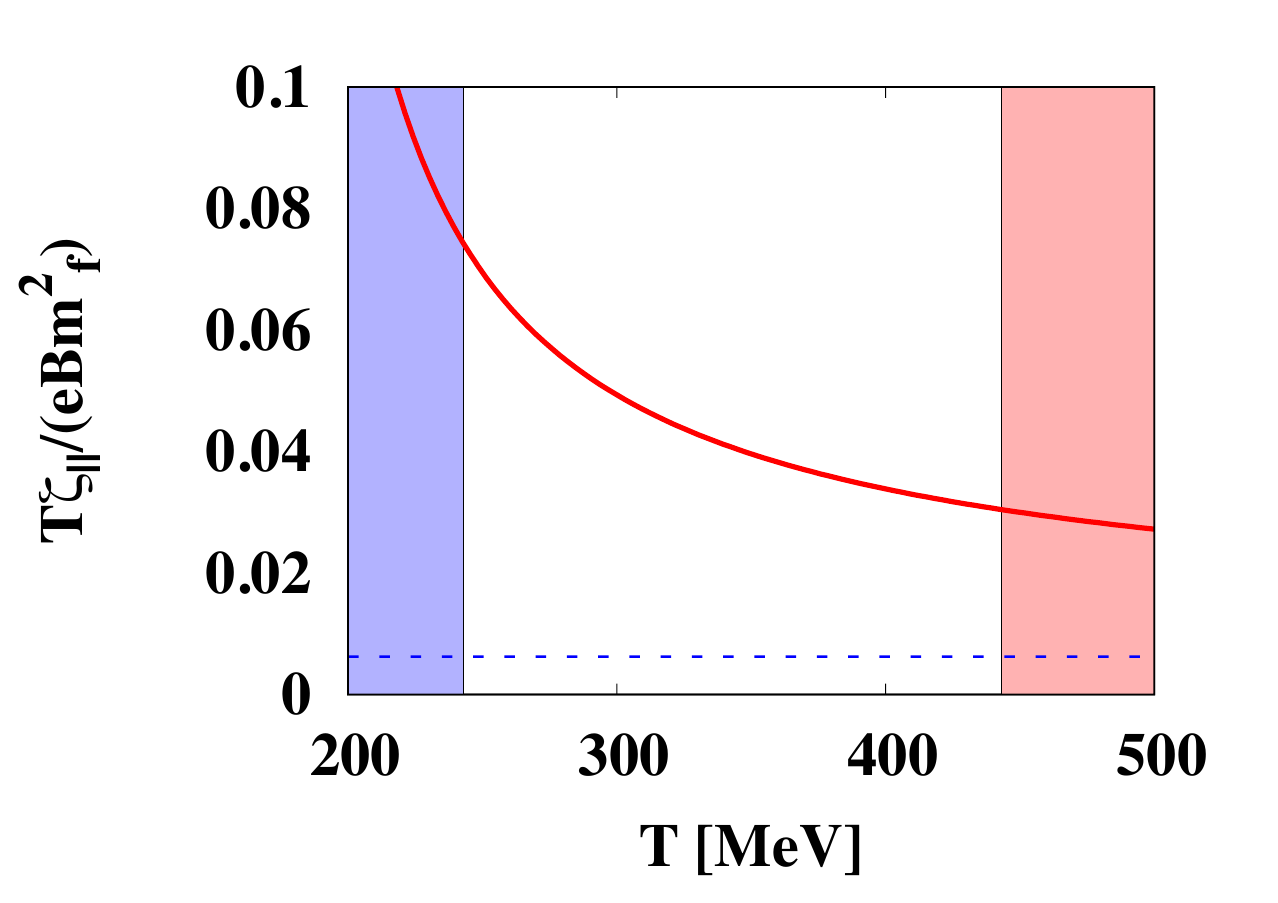}  
\caption{The solid (red) line is our result ($\zeta_\parallel$), and the dotted (blue) line is $\zeta_{B=0}$.
The red (blue) area on the right (left) is the temperature region $\sqrt{eB}<T$ ($T<\sqrt{\alpha_s eB}$) 
in which the LLL (leading-log) approximation is not reliable.
}
\label{fig:bulkviscosity}
\end{center}
\end{figure}

Let us compare this estimate with the one at $B=0$.
The contribution from the $s$ quark is estimated as~\cite{Arnold:2006fz}
\begin{align}
\label{eq:zeta2}
\zeta_{B=0}
&\simeq 0.011 \frac{\mf^4}{\alpha^2_s T}
\simeq 0.12\frac{\mf^4}{T}
,
\end{align}
where we used the same parameters  (\ref{eq:parameter}).
Equations~(\ref{eq:zeta1}) and (\ref{eq:zeta2}) are plotted as functions of $T$ in Fig.~\ref{fig:bulkviscosity}.
To show the limit of our LLL and leading-log approximations, we have colored the temperature regions,
$\sqrt{eB}<T$ and $T<\sqrt{\alpha_s eB}$, in which the two approximations are not justified.
The two approximations are reliable only in the window between these areas. This plot suggests that 
the presence of a strong magnetic field enhances the longitudinal component of the bulk viscosity in 
a wide temperature range. This can actually be understood by looking at the parametric behavior:
Neglecting log factors, we have $\zeta_\parallel\sim eB\mf^2/(g^2T)$ and $\zeta_{B=0}\sim \mf^4/(g^4T)$, and 
their ratio is $\zeta_\parallel/\zeta_{B=0} \sim g^2 eB/\mf^2 \sim (M_g/\mf)^2 $.
From the values estimated around Eq.~(\ref{eq:parameter}), 
this ratio is larger than one with the current values of the parameters.
The temperature dependence in Fig.~\ref{fig:bulkviscosity} comes from
the logarithmic factor in $ \zeta_\parallel $,
because of the temperature-independent infrared cutoff.

\section{Gluon contribution to shear and bulk viscosities}
\label{sec:gluon}
In this section, we estimate the order of magnitude of the contribution of gluons 
to the shear and bulk viscosities. 
All estimates are performed in the LLL approximation assuming that $ |B_f| \gg T^2 $. 
There is no gluon contribution to the conductivities, since the gluon does not carry electric charge, so we do not 
discuss them here.

\subsection{Shear viscosities}

We begin with the five shear viscosities, $\eta_{0,\ldots,4}$.
In this case, the order-of-magnitude estimate can be performed using the schematic expression,
\begin{align}
\eta_i\sim\frac{\text{(typical momentum)}^4} {\text{(mean free path)}^{-1}},
\end{align}
which can be finite without any conformal breaking factor, in contrast to the bulk viscosity.
The typical momentum is apparently of the order of $T$, while the mean-free path needs further consideration.
Let us consider the following three scattering processes:
1) 1-to-2 scattering, 2) 2-to-2 gluon-quark $t$-channel scattering, 
and 3) 2-to-2 gluon-gluon $t$-channel scattering.
They were estimated in Sec.~V and Appendix~B of Ref.~\cite{Hattori:2016lqx} 
in the context of the color conductivity.

It was discussed that the process 1) gives a gluon damping rate ($\sim$inverse of the mean free path), 
which is of the order of $g^2\mf^2eB/T^3$. 
A naive estimate of the contribution from the process 2) to the damping rate is of 
the order of $g^4T eB/\Lambda^2_{\text{IR}}$, 
where the dominant infrared (IR) cutoff $\Lambda^2_{\text{IR}}\sim (g^2eB/\mf^2)^{\frac{2}{3}}T^2$ 
{arises from Landau damping [see discussions below Eq.~(B.9) of Ref.~\cite{Hattori:2016lqx}]. 
The color randomization can be achieved without a momentum exchange, 
so that it is not suppressed in the IR regime. 
However, for the randomization of the gluon momentum, 
which is relevant for the computation of 
viscosities, 
one finds a smaller IR enhancement, 
because the difference between the thermal distribution functions 
in the initial and final states also vanishes in the numerator of the collision integral 
as the momentum transfer decreases. 
An appropriate treatment of this point gives a modification of the parametric estimate} 
as $g^4T eB/\Lambda^2_{\text{IR}}\times(\Lambda_{\text{IR}}/T)^2\sim g^4 eB/T$ (up to a possible logarithmic factor), as was discussed in Appendix~B of Ref.~\cite{Hattori:2016lqx}. 
In the same way, the contribution from 3) is expected to be 
$g^4T^3/\Lambda_{\rm IR}^2\times (\Lambda_{\rm IR}/T)^2\sim g^4T$ 
{with $ \Lambda_{\rm IR} \sim gT $}.
This is smaller than the contribution from 2), so we do not need to take this contribution into account. 

The relative magnitude of the contributions 1) and 2) depends on that of  $\mf $ and  $gT$. 
But actually, whichever the larger contribution is, 
the contribution 2) determines the shear viscosity\footnote{This point was not 
correctly considered in Ref.~\cite{Hattori:2016lqx}.}, as can be seen in the following:
The {Coulomb-gauge} gluon propagator was shown to 
have two orthogonal components around the mass-shell ($p^2\simeq 0$)~\cite{Hattori:2017xoo}:
\begin{align}
\label{eq:gluon-propagator-onshell}
D^{\mu\nu}&\simeq
-\frac{P^{\mu\nu}_T-P^{\mu\nu}_\perp}{p^2-(\varPi_T+\varPi_\parallel)}
-\frac{P^{\mu\nu}_\perp}{p^2-(\varPi_T+\varPi_\perp)}.
\end{align}
The first term has a damping rate given by ${\text {Im}}(\varPi_T+\varPi_\parallel)$ while the other term has 
one determined by ${\text {Im}}(\varPi_T+\varPi_\perp)$.
Here, $\varPi_i$ are the coefficients of the tensor decomposition of the 
retarded gluon self-energy $\varPi^{\mu\nu}$, namely $\varPi^{\mu\nu}=\sum_{i=T,L,\parallel,\perp} \varPi_i P^{\mu\nu}_i$. 
We refer to Ref.~\cite{Hattori:2017xoo} for the definitions of the projection tensors $P^{\mu\nu}_i$. 

The scattering process 1) only contributes to Im$\varPi_\parallel$~\cite{Fukushima:2015wck}, 
while we can show by explicit calculation that 2) contributes to Im$\varPi_T$. 
The perpendicular component $ \varPi_\perp $ is absent. 
Therefore, when the former contribution is much larger than the latter one (Im$\varPi_\parallel \gg $Im$\varPi_T$), 
the contribution to the shear viscosity from the second mode 
in Eq.~(\ref{eq:gluon-propagator-onshell}) is much larger than that from the first mode. 
In the opposite case (Im$\varPi_\parallel \ll $Im$\varPi_T$), 
the damping rates are determined by Im$\varPi_T$ and 
both terms in Eq.~(\ref{eq:gluon-propagator-onshell}) contribute 
to a similar order of magnitude. 

Thus, the inverse of the mean-free path is of order $g^4eB/T$.
Combining these order-of-magnitude estimates, we have 
\begin{align}
\eta_i
&\sim T^4\times \frac{T}{g^4eB}
= \frac{T^5}{g^4eB}.
\end{align}
This is suppressed compared to the value at $B=0$, $\eta\sim T^3/g^4$, 
by a factor of $T^2/eB$.\footnote{
If the contribution of the second term in Eq.~(\ref{eq:gluon-propagator-onshell}) 
vanishes due to the tensor structure or any other reason, 
the shear viscosities will be even more suppressed with only the contribution from the first term, 
when Im$\varPi_\parallel \gg $Im$\varPi_T$.
} 
Physically, this originates from the fact that 
{the gluon damping rate is enhanced by 
the abundance of the quark scatterers, the density of which  
increases   with $eB  $ in the transverse plane}. 

\subsection{Bulk viscosities}

The order-of-magnitude estimate of the gluon contribution to the bulk viscosity is more complicated. 

Let us start with the Kubo formula for $\zeta_\parallel$, Eq.~(\ref{eq:Kuboformula-zeta-para}), 
and see how one can recover the expression in the absence of the magnetic field. 
At $B=0$, the expressions of the two components of the pressure in terms of $T^{\mu\nu}$ reduce to
\begin{align}
\tilde{P}_\parallel &= T^{33}-\Theta_\beta T^{00},\\
\tilde{P}_\perp &= \frac{1}{2}\left(T^{11}+T^{22} \right)
-\Theta_\beta T^{00},
\end{align}
so that $2\tilde{P}_\perp+\tilde{P}_\parallel= \sum_i T^{ii}-3\Theta_\beta T^{00}$.
In this case, we note that $\Theta_\beta=1/3-X$, where $X\sim g^4$ or $m^2/T^2$.
The former contribution to $X$ comes from the conformal anomaly, while the latter one is from the 
explicit breaking of the conformal symmetry by the current quark mass.
Then, Eq.~(\ref{eq:Kuboformula-zeta-para}) gives 
\begin{align}
\begin{split}
\zeta_\parallel 
&= \frac{1}{3} \frac{\partial}{\partial\omega}
{\text{Im}} G^R_{ (2\tilde{P}_\perp+\tilde{P}_\parallel) \tilde{P}_\parallel}
(\vp=\vzero, \omega\rightarrow 0) \\
&=\frac{1}{3} \frac{\beta}{2} 
\Bigl\langle (-T^{\mu}_{\mu} +3X T^{00})_2 \\
&~~~\times \left(T^{33}-\frac{1}{3}T^{00}+XT^{00}\right)_1\Bigr\rangle (p=0) .
\end{split}
\end{align}
The first factor ($-T^{\mu}_{\mu} +3X T^{00}$) vanishes in the conformal case, 
so it yields one conformal breaking factor.
In the other factor $(T^{33}-\frac{1}{3}T^{00}+XT^{00})$, one can replace 
$ T^{33} $ by $ \frac{1}{3} \sum_i T^{ii}  $ because of rotation symmetry 
in the absence of a magnetic field. 
Therefore, this factor also yields the same conformal breaking factor. 
In total, the bulk viscosity at $B=0  $ contains two conformal breaking factors 
as elaborated in Ref.~\cite{Arnold:2006fz, Moore:2008ws}, 
resulting in the parametric estimate shown in Eq.~(\ref{eq:zeta-B=0-schematic}). 
The evaluation of $\zeta_\perp$ can be done in the same way. 
Here, we could recover the expression at $ B=0 $ because of the rotation symmetry 
and the three-dimensional equation of state (EoS), 
$ P = \frac{1}{3} \epsilon + o(m_f^2)+ o (g^2) $. 

Now, let us check what happens in the presence of strong magnetic fields.
In the same way as above, we have 
\begin{align}
\begin{split}
\zeta_\parallel 
&= \frac{1}{3} \frac{\beta}{2} 
\left\langle (-T^{\mu}_{\mu} +3X T^{00})_2
\left(T^{33}-T^{00}+XT^{00}\right)_1\right\rangle (p=0) .
\end{split}
\label{eq:eq1}
\end{align}
Notice the difference to the $B=0$ case: 
Namely, the factor in front of $T^{00}$ in the second factor is unity instead of $1/3$. 
This factor of unity originates from the one-dimensional 
EoS in a strong magnetic field, $ P_\parallel = \epsilon + o (m_f^2)+ o(g^2) $, 
and makes a big difference in the order-of-magnitude estimate, as we will see below.

We focus on the product of $3XT^{00}$ and $T^{33}-T^{00}$ 
taken from the first and second factors in Eq.~(\ref{eq:eq1}), respectively. 
At one-loop order we have
\begin{align}
\label{eq:zeta-parallel-gluon-B=0}
\begin{split}
\zeta_\parallel 
&\sim - \frac{\beta}{2} X
\Bigl\langle \Tc  \left(F^{a0\alpha}F^{a0}{}_{\alpha}\right)_2 \\
&~~~\times \left(-F^{b3\beta}F^{b3}{}_{\beta}+ F^{b0\beta}F^{b0}{}_{\beta}\right)_1
\Bigr\rangle (p=0)\\
&= - \frac{\beta}{2} X
\Bigl\langle \Tc  \left([\partial^0 A^{a\alpha}][\partial^0 A^{a}_{\alpha}] \right)_2 \\
&~~~\times \left(-[\partial^3A^{b\beta}][\partial^3A^{b}_\beta]+ [\partial^0 A^{b\beta}][\partial^0 A^{b}_{\beta}] \right)_1
\Bigr\rangle (p=0), \\
&=  - \beta X (\Nc^2-1)\int\frac{d^4k}{(2\pi)^4} (k^0)^2
\left[(k^0)^2-(k^3)^2\right]\\
&~~~\times \nb(k^0)[1+\nb(k^0)] \rho^{\alpha\beta}(k)\rho_{\alpha\beta}(k) ,
\end{split}
\end{align}
where $\rho^{\mu\nu}(k)$ is the spectral function of the gluon. 
We have used the energy-momentum tensor of the gluons,
\begin{align}
T^{\mu\nu} 
&= \frac{g^{\mu\nu}}{4}F^{a\alpha\beta} F^a_{\alpha\beta}
-F^{a\mu\alpha}F^{a\nu}{}_{\alpha},
\end{align} 
the on-shell condition $k^2\simeq 0$, 
and adopted the Coulomb gauge, in which $A^0$ and $k^iA^i$ do not contribute at this order.
Here, the field-strength tensor is 
$F^{a}_{\mu\nu}\equiv \partial_\mu A^a_\nu-\partial_\nu A^a_\mu -g f^{abc} A^{b}_\mu A^c_\nu$ 
with the $SU(\Nc)$ structure constant $f^{abc}$. 

In the last line of Eq.~(\ref{eq:zeta-parallel-gluon-B=0}), 
we do not see a reason for the factor $(k^0)^2-(k^3)^2$ to vanish. 
This is in contrast to the case of the LLL quark contribution 
which has a (1+1)-dimensional conformal symmetry: 
We have seen that this factor indeed vanishes in the massless limit 
because the LLL quark has a (1+1)-dimensional dispersion relation, 
leaving a small factor of $ m_f^2 $.\footnote{ 
As seen in Eq.~(\ref{eq:disp}), the dispersion relations of the hLLs 
explicitly depend on the field strength $ B $, 
which, thus, will serve as the dominant conformal symmetry breaking factor 
at moderate magnetic field strengths, $ m_f^2 \lesssim eB \lesssim T^2$, 
where populations of thermal excitations in the hLLs are as large as that in the LLL. 
} 
As a result of the mismatch between the dimensions of the quark-dominant EoS and 
of the gluon dispersion relation, the gluon contribution to the bulk viscosities is 
expected to have only one conformal breaking factor. 
Therefore, we get the estimate 
\begin{align}
\begin{split}
\zeta_{\parallel, \perp}
&\sim {\text{(typical momentum)}^4}
\frac{\text{(conformal breaking factor)}} {\text{(mean free path)}^{-1} } \\
&\sim T^4\times \frac{T}{g^4eB} \times X 
\\
&\sim \frac{T^3\mf^2}{g^4eB} ,
\end{split} 
\end{align}
up to a possible logarithmic factor.
When the dimensionless combination $M_g\sqrt{eB}/T^2$ is much larger than unity, 
the gluon contribution is subdominant compared to the quark contribution. 
Also, compared to the bulk viscosity at $B =0  $, 
the gluon contribution is suppressed in a strong magnetic field 
such that $eB \gg T^2 ( T/m_f )^2  $, 
where the ratio $ T/m_f $ is of order one because of the important contribution from $ s $ quarks.

\section{Summary
}
\label{sec:summary}
We have evaluated the longitudinal component of the bulk viscosity $ \zeta_\para $ 
of the QGP in a strong magnetic field, by using the LLL approximation and the Kubo formula.
Together with the longitudinal component of the electrical conductivity~\cite{Hattori:2016cnt, Hattori:2016lqx},
this completes the evaluation of the first-order transport coefficients to which the LLL quarks contribute.
We found that the current quark mass dependence significantly changes compared with the one at 
$B=0$~\cite{Arnold:2006fz}, and explained this behavior as a result of 
the competition between conformal symmetry and chirality conservation.
We also estimated the gluon contribution to the shear and bulk viscosities 
which is suppressed by a large value of the gluon damping rate, which is
enhanced by the density of the LLL quark scatterers $\sim eB  $.

\begin{figure}[t!]
\vspace{-0.7cm}
\begin{center}
\includegraphics[width=0.45\textwidth]{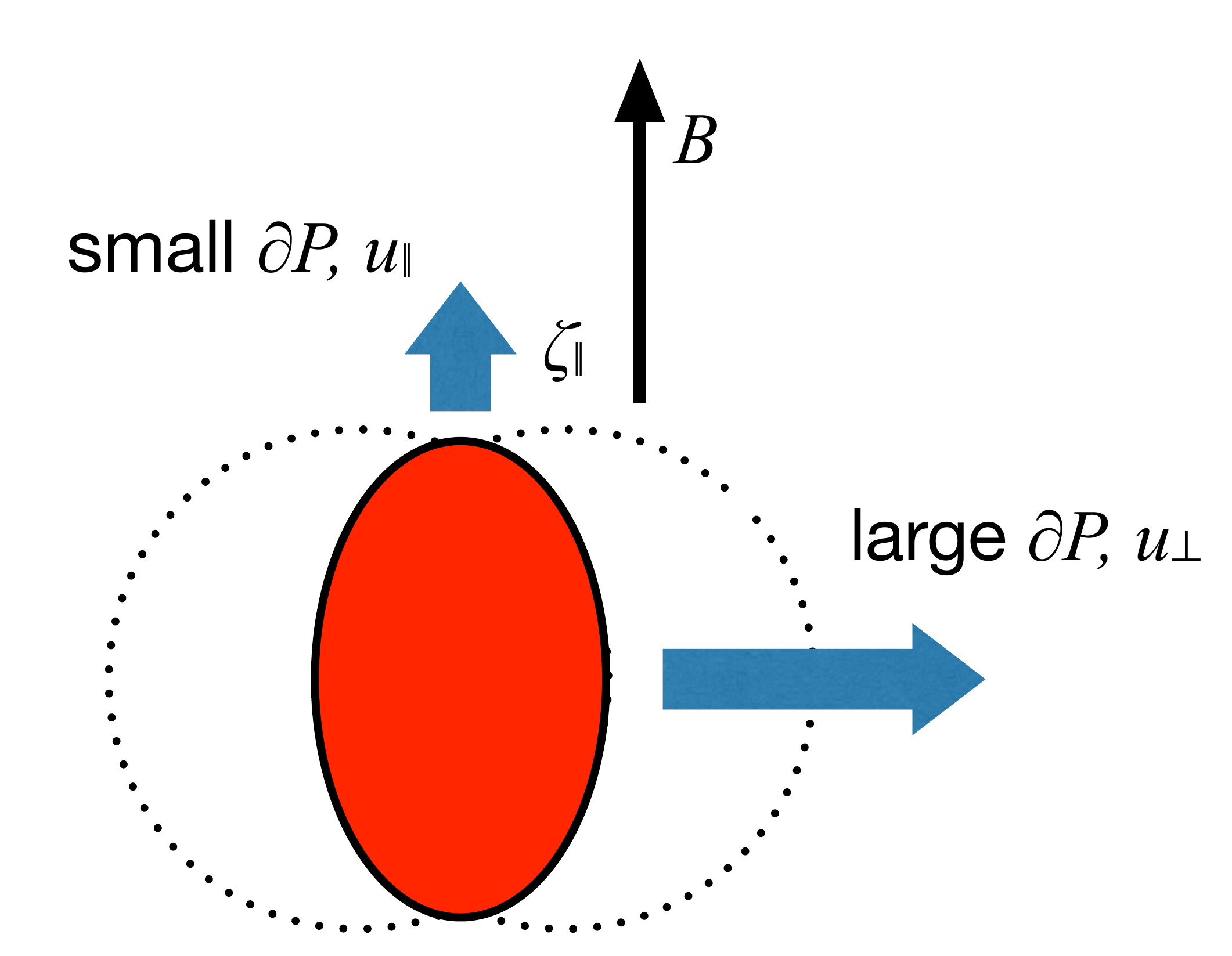}  
\caption{Schematic picture of the elliptic flow and the effect of the anisotropic bulk viscosity on it.
}
\label{fig:flow}
\end{center}
\vspace{-0.5cm}
\end{figure}

When the LLL quark contribution is larger than the gluon contribution, 
the longitudinal component $ \zeta_\para $ is larger than $ \zeta_\perp $. 
This anisotropy may lead to modifications of the hydrodynamic expansion 
of the QGP in heavy-ion collisions. 
Indeed, a large value of  $ \zeta_\para $ has the tendency 
to suppress the hydrodynamic expansion in the direction of the magnetic field, as is schematically sketched in Fig.~\ref{fig:flow}. 
Therefore, our result suggests that a strong magnetic field (which is in general orthogonal
to the reaction plane) potentially 
induces a positive contribution to the elliptic flow measured in heavy-ion collisions. 

To complete the full evaluation of the shear and bulk viscosities in strong magnetic fields 
and to make a more quantitative estimate for phenomenological implications, 
we need to calculate the gluon contributions in more detail. 
We leave this interesting task to future work.

\section*{Acknowledgements}

KH is supported by the China Postdoctoral Science Foundation
under Grant No.~2016M590312 and No.~2017T100266. XGH is supported by the Young 1000 Talents 
Program of China, NSFC with Grant No.~11535012 and No.~11675041, and 
the Scientific Research Foundation of State Education Ministry for Returned Scholars.
DS is supported by the Alexander von Humboldt Foundation.
DHR acknowledges support by the High-End Visiting Expert project GDW20167100136 of the State Administration of Foreign Experts Affairs (SAFEA) of China and by the Deutsche Forschungsgemeinschaft (DFG) through the grant CRC-TR 211 ``Strong-interaction matter under extreme conditions''.

\appendix

\section{Dirac equation in a magnetic field}
\label{app:Dirac-eq}

We briefly summarize the solution for the Dirac equation in a magnetic field:
\begin{eqnarray}
\left( i \slashed D - m_f \right) \psi = 0
\label{eq:Dirac}
\, .
\end{eqnarray}
We introduce the operators $a, a^\dagger = i [D^1 \pm  {\rm sgn}(B_f) i D^2]/\sqrt{2|B_f|} $
where the upper and lower signs are for $  a$ and $a^\dagger  $, respectively. These operators
satisfy $ [ a,  a^\dagger]=1 $.
Then, the Dirac operator is cast into the form
\begin{eqnarray}
i \sla D -m_f  =  i \sla \partial _L -m_f - \sqrt{2|B_f|} \gam^1 (a \prj_+ +a^\dagger \prj_-)
\, .
\nn
\\
\end{eqnarray}
Spin eigenstates $ \psi_\pm=\prj_\pm\psi $ should satisfy
\begin{eqnarray}
 \left[\, \partial_t^2 - \partial_z^2 + (2  a^\dagger a  + 1 \mp 1)  \vert q_f B \vert  + m_f^2 \, \right] \psi_\pm = 0
\, .
\end{eqnarray}
Therefore, the solution for the ground state, the LLL,
is $ \psi_+ = e^{- ip_L \cdot x}  \phi(x_\perp) \prj_+ \chi(p_L)$
with a Dirac spinor $ \chi(p_L) $ and an eigenfunction such that $ a \phi(x_\perp) = 0  $.
We find the LLL dispersion relation to be $ (\eL_p)^2 = (p^3)^2 + m_f^2 $.

Inserting $\psi_+  $ of the LLL into the Dirac equation,
one finds that the Dirac spinor $ \chi(p_L) $
obeys the (1+1)-dimensional ``free'' Dirac equation
\begin{eqnarray}
 (\sla p_L -m_f ) \chi (p_L) = 0
 \label{eq:free}
 \, .
\end{eqnarray}
This equation indicates that the LLL spinor depends only on the longitudinal momentum $ p_L $,
and the LLL fermions have a (1+1)-dimensional ``free'' propagator.
From the condition $a \phi(x_\perp) = 0  $,
the normalized wave function in the Landau gauge is obtained as
\begin{eqnarray}
\phi (x_\perp) = e^{ i p^2 x^2} {\mathcal H} \left(x^1 - \frac{p^2}{B_f}\right)
\, ,
\end{eqnarray}
where
\begin{eqnarray}
{\mathcal H}(x) = \left(\frac{|B_f| } { \pi} \right)^{\frac{1}{4}} e^{- \frac{x^2 }{2} |B_f| }
\, .
\end{eqnarray}
We use the following properties of the Hermite function in the LLL:
\begin{eqnarray}
&&
\int \!\!  dx \, \Ham (x) \Ham (x) = 1\, ,
\\
&&
\int \!\!  dx \, x\, \Ham (x) \Ham (x) =0 \, ,
\\
&&
\int \!\!  dx \, \Ham(x) \partial_x \Ham_n(x) = 0 \, .
\end{eqnarray}

Finally, we count the density of states in a finite box~\cite{Hattori:2015aki, tong2016lectures}.
To this end, note that the second component $ p^2 $ of the canonical momentum
serves as a label of the degenerate states,
and the center coordinate of the cyclotron motion is given by $ x_c^1 =p^2/B_f $.
Accordingly, when the cyclotron center is located
within the length of the system $0 \leq  x_c^1 \leq L_1 $,
we have $ 0 \leq p_y \leq  B_f L_1$ when $\sgn(B_f) >0  $,
and $ - | B_f| L_1 \leq p^2 \leq 0$ when $ \sgn(B_f) <0 $.
Therefore, we get the density of states in the transverse plane as
\begin{eqnarray}
\frac{1}{L_1}  \int_0^{| B_f| L_1} \!\! \frac{dp^2}{2\pi} = \frac{|B_f|}{2\pi}
\, .
\end{eqnarray}


\section{Thermodynamic quantities in the LLL approximation}
\label{app:thermodynamic}

In this Appendix, we first evaluate $\Theta_\beta$, which is required for
the calculation of the bulk viscosity in the LLL approximation.
Before we do so, we need to obtain the energy density and the pressure in the direction of $B$:
The contribution from the quarks in the LLL reads
\begin{align}
\epsilon&= \frac{|\Bf|}{2\pi}\Nc\frac{2}{\pi}
\int^\infty_0 dk^3 \frac{(\eL_k)^2}{\eL_k} \nf(\eL_k)  ,\\
\label{eq:Pparallel}
P_\parallel&=  \frac{|\Bf|}{2\pi}\Nc\frac{2}{\pi}
\int^\infty_0 dk^3 \frac{ (k^3)^2}{\eL_k} \nf(\eL_k) .
\end{align}
In the massless limit, both quantities are equal, as $\epsilon=P_\parallel=|\Bf|\Nc  T^2/12$.
In this case, $\Theta_\beta =1 $, which corresponds to $X=0$.
This result can be understood by remembering that, at $B=0$,
conformal symmetry requires that $\Theta_\beta=1/3$,
since the number of spatial dimensions is three.
In our case, the number of spatial dimensions is effectively reduced to one
thanks to the strong magnetic field, so $\Theta_\beta$ is unity.

As we have seen in Eq.~(\ref{eq:trace}), we need to evaluate the deviation from the $\mf=0$ case.
Therefore, what we should evaluate is
\begin{align}
\epsilon-P_\parallel
&= \mf^2 \frac{|\Bf|}{2\pi}\Nc\frac{2}{\pi}
\int^\infty_0 dk^3 \frac{1}{\eL_k} \nf(\eL_k) .
\end{align}
We note that this integral has a logarithmic infrared divergence in the massless limit.
At the leading-log accuracy in terms of $\ln(T/\mf)$, we get
\begin{align}
\epsilon-P_\parallel
&\simeq \mf^2 \frac{|\Bf|}{2\pi}\Nc\frac{1}{\pi}
\ln\frac{T}{\mf}.
\end{align}
Now, we can obtain $\Theta_\beta$ as
\begin{align}
\label{eq:Theta-beta}
\Theta_\beta
&= 1-\frac{\partial }{\partial \epsilon}(\epsilon-P_\parallel)
\simeq 1- \frac{3 \mf^2}{\pi^2 T^2} ,
\end{align}
in the temperature region $\mf\ll T$.
Here we have used the property $df/d\epsilon=(df/dT)/(d\epsilon/dT)$ for a fixed value of $B$.
This result corresponds to $X=3\mf^2/(\pi^2T^2)$.

Next, we evaluate $\Phi_\beta$. To this end, we need to evaluate $M$ first.
From the definition of $M$ and Eq.~(\ref{eq:Pparallel}), we have 
\begin{align}
\label{eq:magnetixation-LLL}
M= \frac{P_\parallel}{B},
\end{align}
 which is quite different from the expression at weak $B$.
From this expression, we get
\begin{align}
\begin{split}
\Phi_\beta
&= - \left(\frac{\partial P_\para}{\partial \epsilon}\right)_{B}
=-1+ \left(\frac{\partial (\epsilon-P_\para)}{\partial \epsilon}\right)_{B} .
\end{split}
\end{align}
By comparison to Eq.~(\ref{eq:Theta-beta}), we have
\begin{align}
\label{eq:Phibeta-Thetabeta}
\begin{split}
\Phi_\beta
&= -\Theta_\beta .
\end{split}
\end{align}

Finally, let us comment on the interpretation of Eq.~(\ref{eq:magnetixation-LLL}):
In the case of a weak magnetic field, the effect of $B$ on the energy or pressure comes mainly from the Zeeman effect, 
namely the fact that flipping spins of the quarks costs energy.
Therefore, the magnetization is proportional to the total number of spins of the quarks, as $M$ is the pressure difference when we increase $B$. 
In contrast, in the presence of a strong magnetic field, the spins of the quarks in the LLL are always pointing in the direction of $B$ and 
they cannot be flipped.
Instead, the main effect of $B$ on the energy or pressure is to create a large quark density in the transverse plane, due to the 
degeneracy of the LLL.
For this reason, the magnetization in a strong magnetic field $B$ agrees with the pressure due to the quarks, divided by the 
degeneracy of the LLL, as can be see from the above result.

\section{Equivalence to linearized Boltzmann equation}
\label{app:Boltzmann}

In this Appendix, we show that the linearized Boltzmann equation can reproduce the result for the 
bulk viscosity (\ref{eq:result-bulkvis}) obtained from the diagrammatic calculation.
We follow the strategy of Ref.~\cite{Arnold:2006fz}:
We consider the situation that the system is at equilibrium and at rest in the beginning, so that the distribution 
functions are given by the standard Fermi and Bose ones.
Then, the system is disturbed by an expansion in the direction of $B$.
Linear-response theory requires to evaluate the change in the pressure, in order to get information on the bulk viscosity.

The time evolution of the system is described by the Boltzmann equation for the quark distribution function ($f$) in 
the LLL, which is effectively a $(1+1)$-dimensional equation:
\begin{align}
\label{eq:Boltzmann-eq}
(\partial_t +v^3 \partial_z ) f(k^3,t,z)
&= C[f],
\end{align}
where $v^3\equiv k^3/\eL_k$ is the velocity in the direction of $B$, $(t,z)$ are the space-time coordinates, and 
$C[f]$ is the collision term for the 1-to-2 process, the expression of which 
is given in Ref.~\cite{Hattori:2016cnt, Hattori:2016lqx}.

In general, the distribution function can be written in terms of the deviation from the equilibrium value, namely
\begin{align}
f(k^3,t,z)
&=f_{\text{eq}}(k^3,t,z)
+\delta f(k^3,t,z),
\end{align}
where
\begin{align}
f_{\text{eq}}(k^3,t,z)
&\equiv (\exp\{\beta(t)\gamma_u[\eL_k-k^3 u^3(z)] \}+1)^{-1}
\end{align}
is the distribution function in equilibrium in the presence of the flow ($u^3$).
Here $\gamma_u\equiv [1-(u^3)^2]^{-1/2}$ is the gamma factor.
For $u^3=0$, $f_{\text{eq}}$ reduces to $\nf$.
We note that the temperature depends on time, since the expansion decreases the energy density of the system.

We consider the linear response regime, so $ u^3 $ and $\delta f$ are assumed to be small.
Then, the left-hand side of Eq.~(\ref{eq:Boltzmann-eq})
is approximated as
\begin{align}
\begin{split}
&-\nf(\eL_k)[1-\nf(\eL_k)]
\left[\eL_k\partial_t \beta -\beta v^3 k^3 \expansion(z) \right] \\
&= -\beta \nf(\eL_k)[1-\nf(\eL_k)] \expansion(z)
\left(\eL_k \Theta_\beta - v^3 k^3  \right) ,
\end{split}
\end{align}
where $\expansion (z)\equiv \partial_z u^3(z)$ represents the magnitude of the expansion.
In the last line, we have used $\partial_t \beta=\beta \Theta_\beta \expansion(z)$~\cite{Arnold:2006fz}.
The right-hand side of Eq.~(\ref{eq:Boltzmann-eq}) vanishes in equilibrium,
so that it is of linear order in $\delta f$.
In the relaxation-time approximation, the collision term is given by $-\tau^{-1}_k \delta f(k^3,t,z)$
with the parameter $\tau_k$ being the relaxation time.
Combining both sides, the solution of the Boltzmann equation reads
\begin{align}
\delta f
&=  \tau_k \beta \nf(\eL_k)[1-\nf(\eL_k)] \expansion(z)
\left(\eL_k \Theta_\beta - v^3 k^3  \right).
\end{align}
We note that, in the massless limit where the system becomes conformal\footnote{Here,
the conformal symmetry is defined not in $(3+1)$ but in $(1+1)$ dimensions.}
in the classical limit,
the quantity in the bracket vanishes [see the definition of $v^3$ and $\eL_k$, and Eq.~(\ref{eq:Theta-beta})].
This means that the system still persists to be at equilibrium even in the presence of the expansion, 
which is a natural consequence of the conformal invariance.

The bulk viscosity appears in the constitutive relation (\ref{eq:Tmunu-dis}) as
\begin{align}
\label{eq:bulk-P}
\delta P_\parallel
&= -3\zeta_\parallel \expansion,
\end{align}
where $\delta  P_\parallel$ is the deviation of $ P_\parallel$ from the equilibrium value.
Here, we have omitted terms proportional to other transport coefficients
which do not have contributions from the LLL quarks.
Thus, we need to evaluate $\delta  P_\parallel$.
Naively, it is given by
\begin{align}
\label{eq:P-naive}
\delta 
P_\parallel
&= \frac{|\Bf|}{2\pi}\Nc\frac{2}{\pi}
\int^\infty_0 dk^3 \frac{1}{\eL_k}  (k^3)^2
\delta f(k) ,
\end{align}
where we replaced $\nf$ by $\delta f$ in Eq.~(\ref{eq:Pparallel}).
However, we note that, even when $\delta f=0$, the pressure changes
since the temperature decreases in time due to the expansion.
Therefore, we need to subtract this effect, which is found
to require subtraction of $\Theta_\beta T^{00}$ from $T^{33}$~\cite{Arnold:2006fz}.
$T^{00}$ can be expressed in terms of $\delta f$ as in Eq.~(\ref{eq:P-naive})
with the replacement of $(k^3)^2$ by $(\eL_k)^2$.
The subtracted result is found to be
\begin{align}
\begin{split}
\delta \left[
P_\parallel
-\Theta_\beta \epsilon
\right]
&= \frac{|\Bf|}{2\pi}\Nc\frac{2}{\pi}
\int^\infty_0 \!\! dk^3  \frac{\delta f(k)}{\eL_k}
[ (k^3)^2 -\Theta_\beta (\eL_k)^2]
 \\
&= -\frac{|\Bf|}{2\pi}\Nc\frac{2}{\pi} \beta
\int^\infty_0 \!\! dk^3 \frac{[ (k^3)^2 -\Theta_\beta (\eL_k)^2]^2}{(\eL_k)^2}  \\
&~~~\times \tau_k  \nf(\eL_k)[1-\nf(\eL_k)] \expansion(z)
.
\label{eq:bulk-Boltzmann}
\end{split}
\end{align}
We see that another conformal breaking factor $(k^3)^2 -\Theta_\beta (\eL_k)^2$ appears
in addition to the one in Eq.~(\ref{eq:P-naive}).
By identifying $\tau^{-1}_k= 2\xi_k$, we find that the expression for
the bulk viscosity obtained from this equation and Eq.~(\ref{eq:bulk-P})
is identical to Eq.~(\ref{eq:bulkvis-quarkdamp}) from the diagrammatic method.

The equivalence beyond the relaxation-time approximation can also be shown,
as was done in Ref.~\cite{Hattori:2016cnt} in the case of the electrical conductivity.


\bibliography{reference}

\end{document}